\documentclass{article}

\usepackage{PRIMEarxiv}

\usepackage[utf8]{inputenc} 
\usepackage[T1]{fontenc}    
\usepackage{hyperref}       
\usepackage{url}            
\usepackage{booktabs}       
\usepackage{amsfonts}       
\usepackage{microtype}      
\usepackage{fancyhdr}       
\usepackage{graphicx}       
\graphicspath{{media/}}     

\usepackage{subfigure}
\usepackage{caption}
\usepackage{multirow}
\usepackage{enumitem}

\title{Towards Sustainability of Systematic Literature Reviews
}

\author{
  Vinicius dos Santos, Anderson Yoshiaki Iwazaki \\
  University of São Paulo (ICMC-USP) \\
  São Carlos - São Paulo - Brazil\\
  \texttt{\{vinicius.dos.santos, iwazaki.anderson\}@usp.br} \\
   \And
  Katia Romero Felizardo, Érica Ferreira de Souza \\
  Federal University of Technology - Paran\'a (UTFPR) \\
  Cornélio Procópio - Paraná - Brazil \\
  \texttt{\{katiascannavino, ericasouza\}@utfpr.edu.br} \\
  \And
  Elisa Yumi Nakagawa \\
  University of São Paulo (ICMC-USP) \\
  São Carlos - São Paulo - Brazil\\
  \texttt{elisa@icmc.usp.br} \\
}

\begin{document}
\maketitle

\begin{abstract}

\textbf{Background}: The software engineering community has increasingly conducted systematic literature reviews (SLR) as a means to summarize evidence from different studies and bring to light the state of the art of a given research topic. While SLR provide many benefits, they also present several problems with punctual solutions for some of them. However, two main problems still remain: the high time-/effort-consumption nature of SLR and the lack of an effective impact of SLR results in the industry, as initially expected for SLR. \textbf{Aims}: The main goal of this paper is to introduce a new view --- which we name Sustainability of SLR --- on how to deal with SLR aiming at reducing those problems. \textbf{Method}: We analyzed six reference studies published in the last decade to identify, group, and analyze the SLR problems and their interconnections. Based on such analysis, we proposed the view of  Sustainability of SLR that intends to address these problems. \textbf{Results}: The proposed view encompasses three dimensions (social, economic, and technical) that could become SLR more sustainable in the sense that the four major problems and 31 barriers (i.e., possible causes for those problems) that we identified could be mitigated. \textbf{Conclusions:} The view of Sustainability of SLR intends to change the researchers' mindset to mitigate the inherent SLR problems and, as a consequence, achieve sustainable SLR, i.e., those that consume less time/effort to be conducted and updated with useful results for the industry.
\end{abstract}


\section{Introduction}

The community of software engineering (SE) has increasingly adop\-ted Systematic Literature Review (SLR) in recent years \cite{Mendes2020} as a reliable method to summarize evidence from a number of studies and find out the state of the art in a given research topic \cite{Nakagawa17,Kitchenham15}.
SLR has presented many important benefits, including the possibility of dealing with information from different studies in an unbiased manner \cite{Kitchenham15,Niazi15}, producing auditable and repeatable results \cite{Kitchenham2011Repeatability,Budgen2018Reporting}, and identifying research gaps and also perspectives for future investigations \cite{Kitchenham15}.

At the same time, SLR has also suffered from diverse problems. Studies reported some of them, e.g., poor documentation of SLR conducted \cite{Zhou2016Threats,Ampatzoglou2019} and lack of rigor in following well-experimented guidelines for SLR directly impacting the SLR quality \cite{Kuhrmann2017}. Important solutions have been proposed for specific problems, such as techniques for better planning SLR \cite{Cairo2019,Felizardo17b}, conducting SLR 
\cite{Felizardo2017Analysing}, reporting \cite{Cartaxo2018Role}, and adopting supporting tools \cite{Marshall2013}. However, two main problems still remain with no suitable solutions: (i) the SLR conduction is still a very time- and effort-consuming task \cite{Felizardo2020Automating}; and (ii) the lack of effective impact of SLR results in the industry \cite{Badampudi2019Contextualizing}. This current scenario of SLR leads the SE community to question if it is worth continuing to invest in the SLR conduction. Hence, the research question addressed in this work is: Is it possible to reduce the high consumption of resources (time and effort) while assuring to achieve more accessible and useful SLR for the industry? Another question that intrigues us is: Should the SLR researchers change the mindset on how to better deal with SLR conduction and dissemination?

The main goal of this work is to introduce a new view --- which we name Sustainability of SLR --- to better deal with SLR and reduce those two main problems (high resources consumption and lack of impact in industry) and, as a consequence, achieve sustainable SLR (i.e., SLR that consume fewer resources to be conducted and updated with useful results for the industry).
To achieve our goal and answer the research question, we carefully examined six reference studies published in the last decade \cite{Ampatzoglou2019}\cite{Budgen2018Reporting}\cite{Imtiaz2013}\cite{Kitchenham2013}\cite{Riaz2010}\cite{Zhou2016Threats} to identify, group, and systematically analyze the main problems in both SLR conduction and update. 
We found four groups of problems and 31 barriers (i.e., possible causes of the problems). 
Following this, we proposed the view of Sustainability of SLR, which connects these problems among them as well as their causes and encompasses three perspectives (social, economic, and technical) that could somehow deal with the SLR problems. The social perspective addresses human aspects, such as reviewers' communication and stakeholders' participation during the SLR conduction. The economic one is related to the resources (effort and time) to conduct and update an SLR. The technical perspective is related to the supporting tools and technologies used to conduct and update SLR. 

It is worth highlighting that this prospective view does not intend to solve all current problems of SLR; instead, the main contribution of this work is to raise the awareness of researchers to change the mindset when proposing new solutions for SLR problems through a more holistic view.


The remainder of this paper is organized as follows: Section~\ref{sec:research_method} presents the research method. Section~\ref{sec:results} presents the SLR problems identified in the literature and their possible causes; Section~\ref{sec:sus4slr} presents the view of Sustainability of SLR; Section~\ref{sec:discussion} discuss some challenges for achieving the Sustainability of SLR as well as the threats to validity; and Section~\ref{sec:conclusion} presents the final remarks.

\section{Research Method} \label{sec:research_method}

To propose the view of Sustainability of SLR, it was necessary first to survey the literature and understand the current problems of both SLR conduction and update. For this, we selected six reference studies published over the last decade (2010 - 2020) \cite{Ampatzoglou2019,Budgen2018Reporting,Zhou2016Threats,Riaz2010,Kitchenham2013,Imtiaz2013} based on our group experience in conducting SLR. We also used experts' opinions to validate our choice and mainly avoid bias. 

The six studies selected are tertiary studies (SLR of secondary studies) and experience reports from researchers' perspectives to identify the main problems of conducting and updating SLR. Riaz et al. (2010) \cite{Riaz2010} presented an experience report involving three Ph.D. students that identified the main difficulties of novices while they conduct SLR and compared the results with SLR expert experience. Imtiaz et al. (2013) \cite{Imtiaz2013} conducted a tertiary study that evaluated 116 secondary studies published from 2005 to 2011, summarized the experience of researchers, and highlighted the challenges in conducting SLR. Kitchenham and Brereton (2013) \cite{Kitchenham2013} conducted a tertiary study that assessed 68 secondary studies published from 2005 to 2012 to summarize the researchers' opinions about their experiences of conducting SLR and discuss techniques that could be used to improve the SLR process. Bugden et al. (2018) \cite{Budgen2018Reporting} conducted another tertiary study assessing 178 studies published from 2010 to 2015 summarizing lessons learned, researchers' main problems, and focusing on how good SLR could be reported. 

We can also interpret threats to validity as problems in SLR that could not be solved providing clues about the researchers' difficulties during the SLR conduction. Hence, two tertiary studies~\cite{Ampatzoglou2019,Zhou2016Threats} identified and discussed the most common threats to validity and strategies to mitigate them. Zhou et al. (2016) \cite{Zhou2016Threats} assessed 316 studies published from 2004 to 2015 and Ampatzoglou et al. (2019)~\cite{Ampatzoglou2019} analyzed 165 studies published from 2007 to 2016. 
\begin{figure*}[!ht]
    \centering
    \includegraphics[width=1\linewidth]{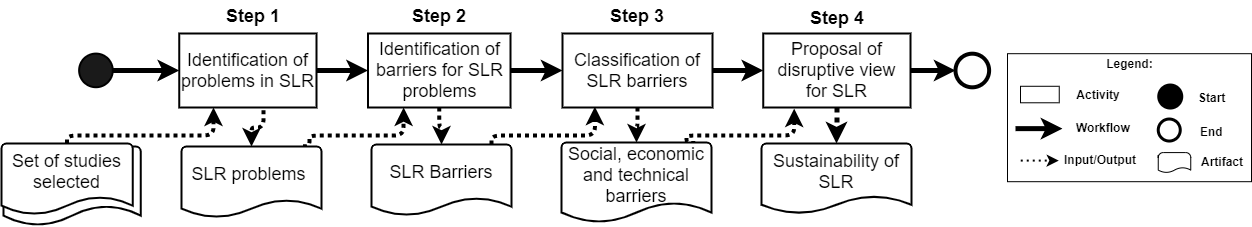}
    \caption{Research method}
    \label{fig:researchMethod}
\end{figure*}


Based on the six studies aforementioned, we performed the four main steps, as shown in Figure \ref{fig:researchMethod}, which summarized the research method adopted. In \textbf{Step 1}, we systematically identified the main problems that occur during the SLR conduction and reported in the six studies. We also deeply checked the studies cited by these six studies. In \textbf{Step 2}, to understand the source of those problems, we carefully identified the current barriers, i.e., the possible causes for these problems, that could be hampering authors to conduct good and effective SLR. In \textbf{Step 3}, we deeply analyzed and summarized those barriers removing duplication and classifying them\footnote{Replication package: \url{https://github.com/CSM-Research/06-LR-TSInSLR}} according to three dimensions (or perspectives): social, economic, and technical. This classification was based on well-known sustainability dimensions already explored in other areas \cite{Becker2015Sustainability, Hill1997, keeble1988brundtland, Purvis2018three, Spangenberg2002, Soini2014}, as described below:

\begin{itemize}[leftmargin=*]

\item \textit{Social barriers} are mainly associated with human aspects of SLR (e.g., communication, culture) and how SLR results can impact the society (academia and industry);

\item \textit{Economic barriers} are directly associated with the high consumption of time and effort while planning, conducting, or updating SLR; and

\item \textit{Technical barriers} address problems with tools and technologies used to support SLR conduction. 

\end{itemize}

It is also worth highlighting that Steps 1, 2, and 3 were performed considering our long-term experience researching, conducting, and updating SLR. Besides, we based on evidence from the literature to make decisions. Finally, in \textbf{Step 4}, we analyzed the SLR problems and proposed a new view that connects these problems and their interactions as well as a possible means to deal with these problems. For this, we inspired ourselves on the views of sustainability from other areas, in particular, the ecology area \cite{Brown1987Global} and SE area~\cite{Becker2015Sustainability}, to derive the view of Sustainability of SLR.


\section{SLR problems and barriers} \label{sec:results}

After conducting Steps 1 to 3 of our research method,  Table~\ref{tab:problemsAndBarriers} summarizes the four main problems were identified: (i) lack of industry's interest in SLR results, (ii) SLR conduction is still very, time and effort consuming (iii) poor documentation of SLR, (iv) lack of SLR's quality. For each large problem, we identified the barriers (i.e., possible causes for these problems and listed in column 4 of Table~\ref{tab:problemsAndBarriers}) and classified them in social, economic, or technical dimensions (shown in column 3). 

\begin{table*}[!ht]

\caption{Problems of SLR and possible causes}
\label{tab:problemsAndBarriers}
\footnotesize
\begin{tabular}{|p{1.6cm}|c|c|p{7.5cm}|p{2.2cm}|}
\hline
\centering \textbf{Problems} & \centering \textbf{ID} & \textbf{Dimension} & \centering \textbf{Barriers (Possible causes)} & \textbf{Reference} \\ \hline
\multirow{4}{*}{\begin{minipage}{1.6cm} \centering Lack of industry's interest in SLR results\end{minipage}} & 1 & Economic/Social & Data synthesis, interpretation, and presentation of results are damaged when researchers have different interpretations of data extracted or when are conducted by a single researcher.  & \cite{Ampatzoglou2019} \cite{Riaz2010} \cite{Zhou2016Threats} \\ \cline{2-5} 
 & 2 & Social & Cultural differences may impact the validity results & \cite{Riaz2010} \cite{Zhou2016Threats} \\ \cline{2-5} 
& 3 & Social & Inappropriate research questions are defined & \cite{Ampatzoglou2019} \cite{Imtiaz2013} \cite{Kitchenham2013} \cite{Riaz2010} \cite{Zhou2016Threats} \\ \cline{2-5} 
& 4 & Social & SLR does not present generalizable results which can be used in industry or other areas & \cite{Ampatzoglou2019}\cite{Budgen2018Reporting} \\ \hline
\multirow{7}{*}{\begin{minipage}{1.6cm} \centering SLR conduction/update is still very time- and effort-consuming \end{minipage}} & 5 & Economic & Lack of standardization and incorrect/incomplete keywords hinder the search string construction & \cite{Ampatzoglou2019}\cite{Imtiaz2013} \cite{Kitchenham2013}\cite{Riaz2010}\cite{Zhou2016Threats}\\ \cline{2-5} 

& 6 & Economic & Inadequate number of studies are evaluated hindering the validity of results & \cite{Riaz2010}
\cite{Zhou2016Threats}  \\ \cline{2-5} 
  & 7 & Economic/Technical & Use of not credible or very specific/broad digital databases may return many irrelevant studies or miss relevant studies &  \cite{Ampatzoglou2019}\cite{Imtiaz2013} \cite{Kitchenham2013}  \cite{Riaz2010} \cite{Zhou2016Threats}\\ \cline{2-5} 
& 8 & Economic/Technical & Primary studies may be duplicated in different databases & \cite{Zhou2016Threats} \\ \cline{2-5} 
& 9 & Economic/Technical & Inaccessibility of resources (papers/databases) & \cite{Zhou2016Threats} \\ \cline{2-5}
& 10 & Economic/Technical & Database limitations and  inefficiencies (e.g., interface, search string syntax) & \cite{Imtiaz2013} \cite{Kitchenham2013}\cite{Riaz2010}  \\ \cline{2-5}
& 11 & Economic/Technical & Need tool to support SLR & \cite{Imtiaz2013}\cite{Kitchenham2013}  \\ \hline 
\multirow{9}{*}{\begin{minipage}{1.6cm} \centering Poor documentation of SLR \end{minipage}} & 12 & Economic & Lack of information about the SLR conduction process (e.g, initial date, duplicated studies). & \cite{Budgen2018Reporting} \cite{Kitchenham2013} \cite{Riaz2010} \\ \cline{2-5} 
 & 13 & Economic & Lack information about the use of inclusion/exclusion criteria. & \cite{Budgen2018Reporting} \cite{Kitchenham2013} \\ \cline{2-5} 
 & 14 & Economic & Insufficient details/report about the quality evaluation & \cite{Budgen2018Reporting} \\ \cline{2-5} 
 & 15 & Economic & Additional search is not being reported correctly & \cite{Budgen2018Reporting} \\ \cline{2-5} 
  & 16 & Economic & Lack of details about the primary studies selected (e.g. context, participants, source material) & \cite{Budgen2018Reporting} \\ \cline{2-5} 
 & 17 & Economic & Lack of information about the exclusion of studies impacting SLR data extraction and synthesis & \cite{Budgen2018Reporting} \\ \cline{2-5} 
 & 18 & Economic & Studies do not specify important SLR details, causing problems to repeatability and replicability & \cite{Ampatzoglou2019}\cite{Riaz2010}\cite{Zhou2016Threats}   \\ \cline{2-5} 
 & 19 & Economic/social & The synthesis process is not clear & \cite{Budgen2018Reporting} \cite{Riaz2010} \\ \cline{2-5} 
 & 20 & Economic/social & Lack information about the reviewers' participation during the SLR conduction process & \cite{Budgen2018Reporting}\cite{Imtiaz2013}\cite{Riaz2010} \\ \hline
\multirow{11}{*}{\begin{minipage}{1.6cm} \centering Lack of quality of SLR \end{minipage}} & 21 & Economic & Many studies do not perform the quality evaluation or do not use it correctly in selection process. & \cite{Budgen2018Reporting} \cite{Riaz2010} \\ \cline{2-5} 
 & 22 & Economic & Few studies had a good search coverage. & \cite{Budgen2018Reporting} \\ \cline{2-5} 
 & 23 & Economic & Inappropriate search methods leading to problems in SLR coverage & \cite{Imtiaz2013}\cite{Zhou2016Threats} \\ \cline{2-5} 
 & 24 & Economic & Selection process can be difficult when inclusion/exclusion criteria are generic or inappropriate. & \cite{Ampatzoglou2019}\cite{Zhou2016Threats}  \\ \cline{2-5} 
 & 25 & Economic & The wrong classification of the primary studies may cause the secondary study to lack robustness. & \cite{Ampatzoglou2019}\cite{Riaz2010}\cite{Zhou2016Threats} \\ \cline{2-5} 
 & 26 & Economic & Delimiting a time span affect the coverage of SLR & \cite{Zhou2016Threats} \\ \cline{2-5} 
 & 27 & Economic & Lack of unpartiality of researchers resulting in bias in study selection & \cite{Zhou2016Threats} \\ \cline{2-5} 
 & 28 & Economic & Extraction process is difficult when quality assessment is biased by reviewer subjectivity or they do not completely understand the data extraction items & \cite{Ampatzoglou2019}\cite{Zhou2016Threats}  \\ \cline{2-5} 
 & 29 & Social & Lack of expert evaluation of the results & \cite{Kitchenham2013} \cite{Riaz2010} \cite{Zhou2016Threats} \\ \cline{2-5}
 & 30 & Economic & Data model and data extraction forms may change during extraction & \cite{Kitchenham2013} \\ \cline{2-5} 
 & 31 & Economic & Difficulties in deciding when to stop the piloting process & \cite{Riaz2010} \\ \hline
\end{tabular}
\end{table*}

For the first problem ``\textit{Lack of industry's interest in SLR results}'', four social barriers were identified and one can be also interpreted as economic. Studies reported that SLR are conducted using inappropriate research questions \cite{Ampatzoglou2019,Imtiaz2013,Kitchenham2013,Riaz2010,Zhou2016Threats} and highlighted that the cultural differences among researchers threats the validity of results (e.g., preferences for the studies of some researchers' nationality) \cite{Zhou2016Threats,Riaz2010}. In addition, the lack of a clear synthesis of SRL results and lack of practical recommendations for SE practitioners impact somehow the generalizability of the results and hamper the use of these results in the industry and academia \cite{Budgen2018Reporting,Ampatzoglou2019}.

For the second problem ``\textit{SLR conduction/update is still very time- and effort-consuming}'', seven barriers were found and all of them are economic barriers, while five of them can also be interpreted as technical barriers. A possible cause for this problem is the lack of standardization of terms to be used in the search string (as the case of SE area \cite{Zhou2016Threats,Ampatzoglou2019,Kitchenham2013,Riaz2010,Imtiaz2013}) and as consequence, the search might return a large number of primary studies (including many irrelevant ones) or miss some relevant studies \cite{Ampatzoglou2019}. Another cause is the inefficiency of electronic databases \cite{Imtiaz2013,Riaz2010,Kitchenham2013} and, despite the improvements in many aspects over the years (e.g., usability, search engine, coverage, interface, etc), several studies still mention problems indicating that the choice of databases is still a challenge~\cite{Ampatzoglou2019}. Five studies mention that non-credible or very specific or broad databases can return many irrelevant studies or can even miss relevant studies, increasing the effort for conducting SLR \cite{Zhou2016Threats,Ampatzoglou2019,Kitchenham2013,Imtiaz2013,Riaz2010}. In this context, many supporting tools were developed to mitigate such problem~\cite{Marshall2015}, especially aiming at reducing the additional effort caused by databases inefficiency or integration problems (e.g., studies indexed by multiple databases). However, many of these tools solve only punctual problems. Hence, the lack of reliable and integrated tools is still considered an important factor for the excessive time and effort consumption.

For the third problem ``\textit{Poor documentation of SLR}'', nine economic barriers were found and two of them can also be interpreted as social barriers. Results indicate a lack of commitment of researchers in following the well-experimented guidelines proposed for reporting SLR \cite{Budgen2018Reporting}. Three studies reported that SLR do not report important details about the conduction process that allows SLR to be updated or replicated \cite{Budgen2018Reporting,Kitchenham2013,Riaz2010}. The poor documentation directly impacts academia and industry by creating a lack of credibility and problems in auditability and reproducibility of SLR \cite{Zhou2016Threats,Ampatzoglou2019,Riaz2010}. In addition, this problem extends to the social dimension since a clear data synthesis is considered very important to disseminate useful results of SLR to industry and also academia \cite{Budgen2018Reporting}. 

For the fourth problem ``\textit{Lack of quality of SLR}'', 10 economic barriers and one social barrier were found. The economic barriers indicate that authors have difficulty in following the guidelines proposed for SLR and ensuring the quality of selection \cite{Zhou2016Threats,Ampatzoglou2019}, data extraction \cite{Riaz2010,Ampatzoglou2019}, synthesis and documentation \cite{Budgen2018Reporting}. The wrong classification of the primary studies or application of inappropriate search methods, for example, cause problems in the SLR coverage and robustness. In addition, the lack of quality can also be considered a social problem because it is recommended that SLR include experts to evaluate the SLR results \cite{Kitchenham2013,Riaz2010,Zhou2016Threats}, because when they are not involved in the SLR process, results cannot be adequately synthesized, even generalized, and useful for SE practitioners.

Observing from a broader perspective, all problems and their possible causes mentioned previously are somehow interconnected among them. For example, using the economic perspective, the poor documentation hampers the auditability of SLR, resulting in a lack of confidence in the SLR results \cite{Zhou2016Threats,Ampatzoglou2019,Riaz2010}. It also hampers the repeatability of SLR; hence, SLR cannot be easily updated \cite{Kitchenham2011Repeatability}, and researchers need sometimes to reconduct the SLR from ``scratch'' and, as a consequence, consuming extra resources (time and effort). This problem is also connected with a social problem because the low quality of SLR also directly impacts the confidence in the results and generates doubts in readers about biases that could be occurred during the SLR conduction. Jointly to the lack of practical recommendations for SE practitioners in such documentation \cite{Badampudi2019Contextualizing,Budgen2018Reporting}, in many cases, SLR are not used source of evidence for decision-making process \cite{Cartaxo2018Role}, which could improve the productivity of industry and academia. Aiming to reduce the time and effort spent on the SLR conduction and deliver fast results, researchers have ignored important well-experimented guidelines \cite{Budgen2018Reporting}, which could ensure such quality, and have created poorly documented SLR with shallow results, which are not useful in the end.

Based on the set of problems, their dimensions, and barriers, we proposed a new view that could suggest to the researchers how to better deal with these problems in an integrated way, as presented in the next section.

\section{Sustainability of Systematic Literature Review}
\label{sec:sus4slr}

Before presenting the view of Sustainability of SLR, we provide a brief overview of the topic of sustainability that has been already researched in other areas and has served as a basis for our work. 

\subsection{Brief Overview of Sustainability}

Sustainability started in the ecology context \cite{Purvis2018three} and became popular in 1987 when the UN World Commission on Environment and Development published the report ``Our Common Future'' (Brundtland Report) defining ``sustainable development'' as a development that meets the needs of the present without compromising the ability of future generations to meet their own needs \cite{keeble1988brundtland}.  

Next, computer science researchers realized that sustainability could be adapted to other contexts, such as SE \cite{Hilty2011Sustainability}. In 2009, the International Conference on Software Engineering (ICSE) started a special track called ``software engineering for the planet''\footnote{\url{https://www.cs.uoregon.edu/events/icse2009/specialSessions/\#planet}} discussing how software can interact with sustainability, including energy consumption and the impact of software production in society \cite{Hilty2011Sustainability}. In the next years, the term ``Sustainable Software'' was interpreted in two ways \cite{Penzenstadler2014Systematic}: (i) the software code being sustainable; and (ii) the purpose of the software being to support sustainability goals, i.e., to improve the sustainability of humanity on our planet. Both interpretations must coincide in a software system that contributes to a more sustainable life, i.e., sustainable software is energy efficient, minimizes the environmental impact of the processes it supports and has a positive impact on social sustainability and/or economics. In 2015, the \textit{Karlskrona manifesto}\footnote{\url{http://www.sustainabilitydesign.org/}} was signed \cite{Becker2015Sustainability}, creating awareness of the SE community on the need for sustainability and becoming a non-functional requirement for software \cite{Raisian2016Current} and aiming to support software engineers to reduce unnecessary resource consumption. 

To adequately address sustainability, it is usually broken down in different dimensions \cite{Purvis2018three}. The Brundtland Report \cite{keeble1988brundtland} described sustainability in ecology with three dimensions: social, economical, and environmental. Although these dimensions have become mainstream throughout the literature, they are not universal \cite{Purvis2018three}. Some authors consider necessary to have additional dimensions (e.g., institutional \cite{Spangenberg2002}, cultural \cite{Soini2014}, or technical \cite{Hill1997}). In the SE community, the Karlskrona Manifesto included the individual and technical dimensions to maintain human capital and ensure the longevity of information, systems, and infrastructure and their adequate evolution with changing surrounding conditions. 

In summary, software construction has a well-defined process, consumes time and effort, uses human resources, and its success depends on efficient tools, human skills, and social aspects (e.g., participation of stakeholders). In this perspective, researchers in the SE area already understood that this problem must be solved using a systemic solution \cite{Becker2015Sustainability}. Coming from this context, we have used all knowledge previously accumulated and well-experimented in the environmental and SE area to propose a view of sustainability that could be valid to SLR.

\subsection{Sustainability of SLR and their Dimensions}





Considering that the problems and barriers found in this work are associated with economic, social, and technical perspectives, we broke down the Sustainability of SLR into these three dimensions, as described below: 

\begin{itemize}[leftmargin=*]

\item \textbf{Social dimension:} it addresses the human aspects associated with SLR, such as communication and collaboration, and ensures effective participation of stakeholders in the SLR process. This dimension focuses on ensuring that current and future stakeholders (including both, academic community and SE practitioners) have open access to the SLR results and ensuring that these results could be easily used.

\item \textbf{Economic dimension:} it preserves the resources (time and effort) during the conduction, audition, and update or replication of SLR by creating high-quality SLR, i.e., this dimension ensures high-quality documentation, credible databases, efficient search engines, and rigor in following the guidelines to preserve reproducibility and auditability, minimizing unnecessary efforts. 

\item \textbf{Technical dimension:} it preserves the reliability of supporting tools used to conduct SLR, facilitating their use for the update or reconduction of those SLR. This dimension must assure technical ways to minimize the efforts by automating tasks and supporting their execution.

\end{itemize}

Based on these three dimensions, the view of Sustainability of SLR can be understood as \textit{``the process and a set of actions to make it possible to preserve SLR that endure over time (i.e., longevity) with less possible time and effort consumed and an effective impact to the industry.''}

Regarding other sustainability dimensions, we believe they are not directly related to SLR. For instance, the environmental dimension handles the impacts of human activities in natural resources (e.g., water, land, air) or, in SE, it manages the excessive amount of energy consumption caused by software. We did not include such a dimension because there is no clear connection between SLR and the environment. Therefore, for the while, we believe that the three dimensions seem to be sufficient to connect SLR problems and understand their connections and possible ways to achieve sustainable SLR.

Following this, we develop a discussion about how to better represent graphically the Sustainability of SLR. For this, we revisit studies in the literature from other areas \cite{Giddings2002Environment,keeble1988brundtland, Purvis2018three} and observed that the sustainability dimensions are described interconnected among them \cite{Purvis2018three}. Therefore, an important discussion in the sustainability context is how to make a fair graphical representation including all dimensions defined and how they interact with each other. The Brundtland Report \cite{keeble1988brundtland} proposed a representation (i.e., a Venn diagram with interlaced circles and the intersection between them) that became the most popular one for sustainability in the ecological area. The key idea is that human society is only sustainable if it can be sustained in all dimensions. 

\begin{figure}
    \centering
    \includegraphics[width=0.85\linewidth]{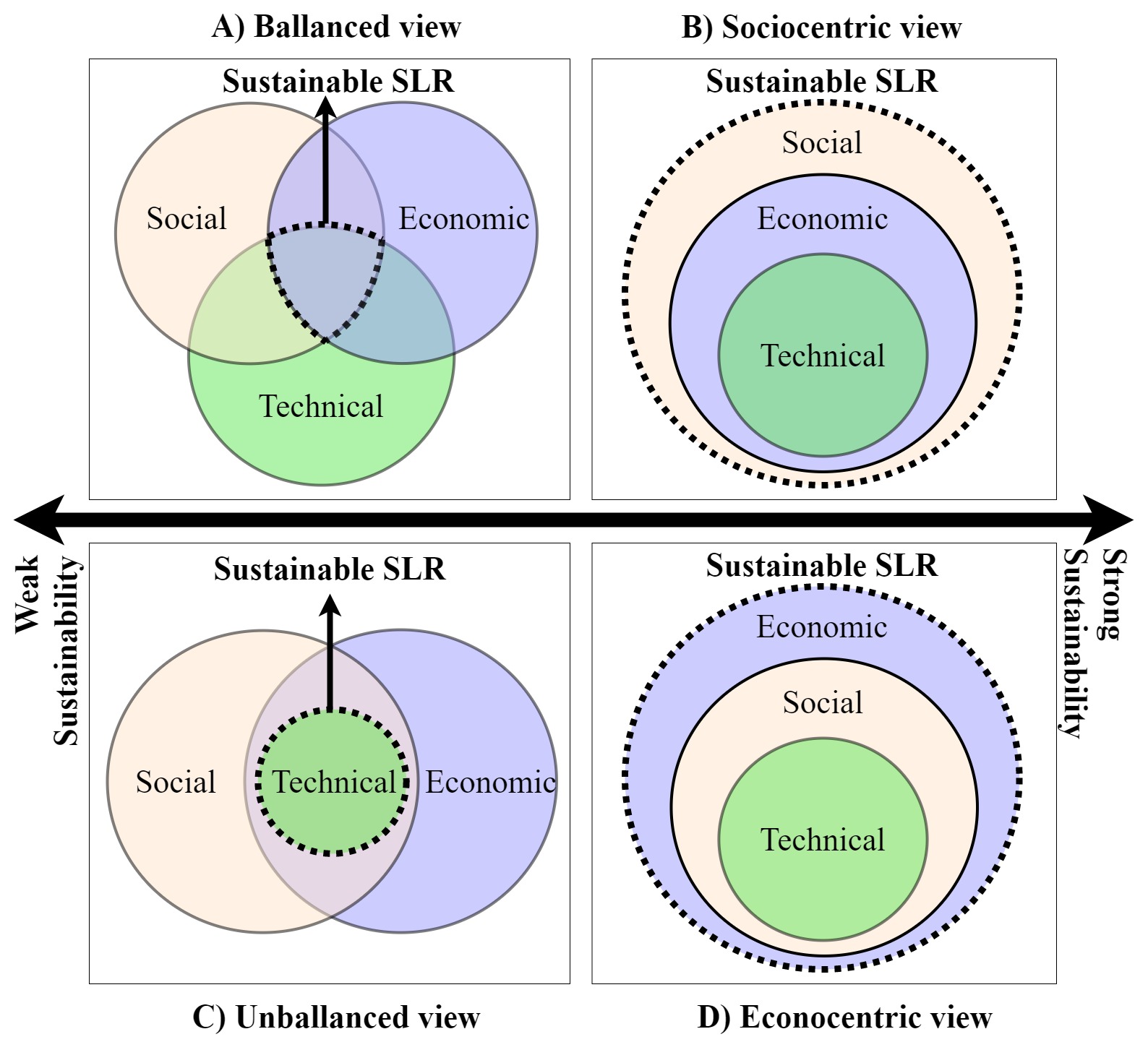}
    \caption{An View of Sustainability of SLR}
    \label{fig:sustainabilityModels}
\end{figure}

Figure \ref{fig:sustainabilityModels} shows a graphical representation of Sustainability of SLR using the Venn diagram. Figure \ref{fig:sustainabilityModels}.A  represents the \textit{Ballanced View} with an intersection of all dimensions. We can observe that attributing the same weight for all dimensions and balancing them to achieve sustainability may lead to questions about its feasibility, due to the large number of factors that impact each dimension. 
An alternative is the \textit{Unballanced View} that considers the technical dimension as a common concern that cannot be dissociated from social and economic perspective, as shown in Figure \ref{fig:sustainabilityModels}.C.
We consider that both representations (shown in Figures \ref{fig:sustainabilityModels}.A and \ref{fig:sustainabilityModels}.C) present a ``Weak Sustainability'' (which was previously defined in \cite{CabezaGuts1996}) because solving all problems associated only with, for instance, social dimension does not assure to achieve a sustainable SLR.

In an opposite perspective, Giddings et al \cite{Giddings2002Environment} proposed a different representation that considers that dimensions must be nested and with different weights. This representation assumes that it is not possible for the society and economy to develop outside the biosphere~\cite{MorandnAhuerma2019} and this perspective warns that a finite planet cannot sustain human life with an economy that intends unlimited growth. It is necessary to acknowledge that there are fundamental biophysical limits that constrain the natural resources on the planet \cite{Neumayer2013}. This perspective arranges the sustainability dimensions providing a hierarchic view, i.e., problems of each dimension must be treated by different disciplines and each decision must consider the constraints of each dimension \cite{Becker2015Sustainability} and this representation was later called ``Strong Sustainability'' \cite{MorandnAhuerma2019} .  

Figure \ref{fig:sustainabilityModels}.B presents the \textit{Sociocentric View} that nests the dimensions considering 
the social dimension is the broadest and technical one is the most specific, so attributing different weights for them. This representation assumes 
technical problems are essentially connected with economic problems, and the social constraints must guide the decision-making of researchers while conducting SLR. Despite this representation seems to be fair for SLR, it may be a problem to put social priorities higher than economic needs. For example, the software industry (which is tightly associated with the social dimension) needs to speed up the SLR conduction; however, it may cause a lack of rigor in documentation and reliability that could directly impact the feasibility and value the of SLR and, as a consequence, it may result in unsustainable SLR. Figure \ref{fig:sustainabilityModels}.D presents the \textit{Econocentric View} that considers that social needs are constrained by economic aspects. Due to the resources constraints sometimes imposed by the economic dimension, researchers are prone to conduct SLR ignoring the social concerns. An example is not to include a given search question that could indeed answer the doubt from a wider community only to speed up the SLR process.

In the end, the graphical representations of Sustainability of SLR intend to provide us the opportunity for better understanding the inherent difficulties to create hierarchies among the dimensions and to better balance them. Considering the current scenario of SLR, we believe that it is necessary to move backward and recognize that the problems cannot be solved without integrated thinking. 

\section{Discussion}\label{sec:discussion}

We believe that this paper provides a good starting point and raises awareness of the community about the current problems of SLR from a holistic panorama. Many of the barriers have been mentioned since 2010 but 
not 
solved 
yet, in particular, the need for adequate synthesis of SLR results to be suitable for the industry. 
Our work indicates 
there is also a lack of: (i) knowledge about the SLR conduction process; (ii) application of the good practices
for SLR conduction; (iii) details and important information to reproduce and update SLR; and (iv) tools' integration. 
We also observed 
many barriers found are related to the conduction process, while few social barriers are reported indicating a lack of knowledge about social problems.

We believe this paper is provocative in the sense that it could bring a disruptive view and, as a consequence, new and coordinated actions, and solutions to make SLR reduce the high consumption of resources while producing useful results for the industry and academia. In this context, create effective methods to include the stakeholders (i.e., members of the industry) could be one of the possible solutions for social dimension problems. Economic dimension must create feasible methods to support the production of high-quality SLR consuming fewer resources (time and effort) to the conduction and update. For this, it is necessary to overcome many challenges including the researcher's mindset about the importance of correctly documenting SLR using the best practices available for reporting and include SLR sustainability as a quality criterion. The technical dimension must support tools and technologies used to plan/conduct/report aiming to ensure its quality, accessibility, and integration. Furthermore, this dimension must handle the slow and fragmented development of tools to support SLR and propose solutions towards integration. For this, it is necessary an international collaboration to provide long-term support and integration of SLR tools. Finally, the main challenge of Sustainability of SLR is to change the researchers' mindset about the importance of considering this integrated and holistic view while conducting SLR and proposing new solutions to overcome the current problems.


Regarding the threats to the validity, 
we did not perform an exhaustive search for studies that could have reported other problems and barriers in SLR. 
It is possible 
some problems and barriers have been missed. Besides, our work considered a 10-year period (2010-2020); hence, it is possible that problems and barriers reported before this period have not been considered. To mitigate these two threats, we used the experts' opinions to validate the set of studies considered in our analysis. Moreover, the results presented in this work could have been influenced by our knowledge and experience in researching and conducting SLR. To mitigate this threat, we strictly based on evidence from the literature to make any decisions.

\section{Final Remarks}
\label{sec:conclusion}


The academic community has increasingly conducted SLR, but they still suffer from critical problems, and the several punctual solutions being proposed for those problems have not been widely effective. This paper is placed in this context and intends to somehow be provocative to call the attention of this community to change its mindset regarding the conduction and use of SLR results. 
Derived from the sustainability concept from other areas, the Sustainability of SLR primarily intends to achieve sustainable SLR, i.e., low consumption of resources (time and effort) and with useful results for the industry. As future work, the community needs to work together to promote sustainable SLR and, as a consequence, to achieve the initial goal of SLR that is the support of the industry.


\section{Acknowledgements}
This study was supported by 
FAPESP (grants: 2015/24144-7, 2019/23663-1),
Capes (grants:  PROEX-11308091/D, PROEX-11357580/D), and CNPq (grant: 312634/2018-8).

 \bibliographystyle{unsrt}  
 \bibliography{references}

\end{document}